
%
\input harvmac
\noblackbox
\Title{UCLA/94/TEP/23}{\vbox{\centerline{Operator Product Expansions in the}
\vskip2pt\centerline{Two-Dimensional O(N) Non-Linear Sigma
Model$^\star$}}}
\footnote{}{$^\star$ This work was supported in part by the U.S. Department
of Energy, under Contract DE-AT03-88ER 40384 Mod A006 Task C.}

\centerline{Hidenori SONODA$^\dagger$\footnote{}{$^\dagger$
sonoda@physics.ucla.edu} and Wang-Chang SU$^*$\footnote{}{$^*$
suw@physics.ucla.edu}}
\bigskip\centerline{\it Department of Physics,
UCLA, Los Angeles, CA 90024--1547, USA}

\vskip 1in
The short-distance singularity of the product of
a composite scalar field that deforms a field theory
and an arbitrary composite field can be expressed geometrically
by the beta functions, anomalous dimensions, and
a connection on the theory space.  Using this
relation, we compute the connection
perturbatively for the O(N)
non-linear sigma model in two dimensions.  We show
that the connection becomes free of singularities
at zero temperature only if
we normalize the composite fields so that
their correlation functions have well-defined
limits at zero temperature.

\Date{June 1994}

\def\O{{\cal O}}
\def\vol{{\rm vol}(S^{D-1})}
\def\p{\partial}
\def\L{{\cal L}}
\def\ep{\epsilon}
\def\dt{{d \over dt}~}
\def\one{{\bf 1}}
\def\vev#1{\left\langle #1 \right\rangle}
\def\vphi{\vec{\phi}}
\def\bra#1{\left[#1\right]}
\def\phione{{\vphi^2 \over 2}}
\def\phitwo{{(\vphi^2)^2 \over 8}}
\def\dphi{{1 \over 2} \p \vphi \cdot \p \vphi}
\def\ddphi{{1 \over 2} (\vphi \cdot \p_\mu \vphi)^2}
\def\g{\tilde{g}}
\def\m{\tilde{m}}

\newsec{Introduction}

In refs.~\ref\rsonodai{H.~Sonoda, Nucl. Phys. {\bf B383}(1992)173},
\ref\rsonodaii{H.~Sonoda, Nucl. Phys. {\bf B394}(1993)302}, and
\ref\rsonodaiii{H.~Sonoda, ``Connection on the theory space,''
a talk given at Strings 93 conference in Berkeley, May 1993, UCLA/93/TEP/21
(hep-th/9306119)} it was found that the singularities of
a conjugate field that deforms a field theory and an arbitrary
composite field admit a geometrical expression.  This paper is a continuation
of the study of the geometrical structure.  The simple example
of the four-dimensional $\phi^4$ theory has been examined in ref.~\rsonodaii.
In the present paper we will study a more non-trivial example of the
two-dimensional non-linear sigma model in some detail.

Let us briefly summarize the geometrical structure obtained in
refs.~\rsonodai, \rsonodaii, and \rsonodaiii.
We consider a finite dimensional theory space
with local coordinates $g^i (i=1,...,N)$, which are nothing
but the parameters of a renormalized euclidean field theory
in $D$-dimensions.  Let the scalar field conjugate
to $g^i$ be $\O_i$.   The conjugate fields form a basis of the tangent
vector bundle of the theory space.   The linearly independent composite
fields $\{\Phi_a\}_g$ make a basis of an infinite dimensional vector bundle.

We denote the renormalization group (RG) equations of the parameters by
\eqn\ebetageneral{\dt g^i = \beta^i (g) .}
The beta functions $\beta^i$ form a vector field on the theory space.
Eq.~\ebetageneral\ implies that the conjugate fields satisfy
\eqn\eRGconjgeneral{\dt \O_i = \left(D \delta_i^{~j} - {\p \beta^j \over \p
g^i}
\right) \O_j .}
We denote the RG equations of the composite fields by
\eqn\eRcompgeneral{\dt \Phi_a = \Gamma_a^{~b} (g) \Phi_b .}

Let us introduce the operator product expansion (OPE) of
a conjugate field and a composite field:
\eqn\eopegeneral{\O_i (r) \Phi_a (0) =
{1 \over \vol} (C_i)_a^{~b} (r;g) \Phi_b (0) + {\rm o} \left( {1 \over r^D}
\right) ,}
where we take the angular average over $r$, and we keep only the part
which cannot be integrated over in space.  Then, the matrix
\eqn\eh{H_i (g) \equiv C_i (r=1;g)}
is a tensor on the theory space.
In ref.~\rsonodaiii\ the following
geometrical expression for the tensor $H_i$ has been written:
\eqn\eformula{H_i (g) = \p_i \Psi (g) + [c_i (g) , \Psi (g)] + \beta^j (g)
\Omega_{ji} (g) ,}
where we define
\eqn\epsi{\Psi (g) \equiv \Gamma (g) + \beta^i (g) c_i (g) ,}
and
\eqn\ecurv{\Omega_{ij} (g) \equiv \p_i c_j - \p_j c_i + [c_i, c_j] .}
The matrix $c_i (g)$ is a connection on the theory space.  It is a connection,
since it transforms as
\eqn\ectrans{ c_i (g) \to N(g) (\p_i + c_i (g)) (N(g))^{-1} }
under an arbitrary change of basis
\eqn\ebasistrans{\Phi_a \to (N(g))_a^{~b} \Phi_b .}
It is easy to check that $\Psi (g)$ is a tensor.  Eq.~\ecurv\
gives a curvature of the connection $c_i$.  The curvature can be
obtained as the double integral over connected
three-point functions \rsonodaiii:
\eqn\ecurvformula{\eqalign{
&\Omega_{ji} \vev{\Phi}_g
= \int_{1 \ge r} d^D r~{\rm F.P.} \int_{1 \ge r'} d^D r' \cr
&\quad \times \vev{ \left( \O_i (r) \left( \O_j (r') - {1 \over \vol} C_j (r')
\right) -
( r \leftrightarrow r' ) \right) \Phi (0)}_g^c ,\cr}}
where F.P. stands for taking the integrable part with respect to $r$.

The connection $c_i$ first appeared as finite counterterms
in the so-called variational formula \rsonodai.
(In two dimensions a connection had been introduced
by Kutasov \ref\rkutasov{D.~Kutasov,
Nucl. Phys. {\bf B307}(1988)417;
Phys. Lett. {\bf B220}(1989)153}.)  The spatial
integral over the conjugate
field $\O_i$ gives a field theoretic realization of the partial
derivative with respect to the coordinate $g^i$, and
finite counterterms are necessary to compensate the
arbitrariness involved in the short-distance regularization
of the spatial integral.  The geometrical formula
\eformula\ implies that the connection $c_i$ actually controls
the short-distance physics together with such obviously
important quantities like the beta functions and anomalous
dimensions.

The purpose of the paper is to study the connection
$c_i$ in the two-dimensional O(N) non-linear sigma model.
The model is interesting in its similarity to the four-dimensional
non-abelian gauge theories: both are asymptotic free, and
the relevant symmetries are realized non-linearly.
We will find that the behavior of the connection $c_i$
at low temperatures (equivalently, at short distances)
contains information on the zero temperature limit of
the theory.  More specifically, we will find that
the elements of the connection $c_i$ are finite at
zero temperature only if we normalize the composite
fields such that their correlation functions have well-defined
and non-trivial zero temperature limits.

The paper is organized as follows.  In sect.~2 we describe the model.
Since the model is well-known, we will be brief.  In sect.~3
we give the results of perturbative calculations of the coefficients
in the OPE's of the conjugate fields and composite fields.
The details are given in the appendices.
In sect.~4 we determine the elements of the connection
$c_i$ perturbatively.  We find that some elements diverge
at zero temperature.  In sect.~5 we study the low temperature
behavior of the theory to understand the physical meaning
of the divergences encountered in sect.~4.  We conclude
the paper in sect.~6.

\newsec{The model}

We define the theory in $D=2+\epsilon$ dimensions using
dimensional regularization and the minimal subtraction (MS) scheme.
The lagrangian is
\eqn\eL{\L = {1 \over g_0} \left( {1 \over 2 } \partial_\mu \Phi_0^I
\partial_\mu \Phi_0^I -
m_0^2 \Phi_0^N \right) ,}
where $\Phi_0^I (I=1,...,N)$ is the bare field, normalized by
\eqn\enorm{\sum_{I=1}^N \Phi_0^I \Phi_0^I = 1 ,}
and $g_0, m_0^2$ are the bare temperature and magnetic field in the N-th
direction.
The renormalized parameters are
\eqn\egm{ g_0 = Z_g (\ep; g) g , \quad m_0^2 = Z_m (\ep; g) m^2 .}
We will suppress the renormalization scale $\mu$.
In actual calculations we choose a particular value for $\mu$
to simplify the results (see sect.~3).
Let us denote the RG equations for the renormalized
parameters by
\eqn\ebeta{\dt g = \beta_g (g) ,\quad
\dt m^2 = (2 + \beta_m (g)) m^2}
in the limit $\ep \to 0$.

The renormalized O(N) vector field $\Phi^I$ is defined by
\eqn\ePhi{ \Phi_0^I = \sqrt{Z_\Phi (\epsilon; g)} ~\Phi^I ,}
where the renormalization constant $Z_\Phi$ satisfies the relation
\ref\rbrezin{E.~Br\'ezin, J.~Zinn-Justin, and J.~C.~Le Guillou,
Phys. Rev. {\bf D14}(1976)2615}
\eqn\ezrel{Z_m \sqrt{Z_\Phi} = Z_g .}

The fields conjugate to $g, m^2$, denoted by $\O_g, \O_m$,
are defined by
\eqn\econj{\eqalign{\O_g &\equiv {\p \L \over \p g}\Big|_{m^2, \Phi_0}
= - {1 \over g^2} B_0 + {m^2 \over g^2} A_1 \cr
\O_m &\equiv {\p \L \over \p m^2}\Big|_{g, \Phi_0} = - {1 \over g} A_1 ,\cr}}
where we define the renormalized fields $B_0, A_1$ as
\eqn\ebzeroaone{\eqalign{
B_0 = \left[ {1 \over 2} \p_\mu \Phi^I \p_\mu \Phi^I \right]
&\equiv \left( 1 + g {\p \ln Z_g \over \p g} \right) {Z_\Phi \over Z_g}
{1 \over 2} \p_\mu \Phi^I \p_\mu \Phi^I - m^2 g {\p \ln Z_\Phi \over \p g}
{1 \over 2} \Phi^N ,\cr
A_1 &\equiv \Phi^N .\cr}}
(The bracket indicates a renormalized field in the MS scheme.)

The conjugate fields satisfy the RG equations (in the limit
$\ep \to 0$):
\eqn\eRGconj{\eqalign{\dt \O_m &= - \beta_m \O_m ,\cr
\dt \O_g &= (2 - \beta_g') \O_g - m^2 \beta_m' \O_m .\cr}}

We cannot possibly consider all the composite fields.
As examples, we examine the four fields
$A_1$, $A_2$, $\p^2 A_1$, and $B_0$,
where we define the (N,N)-component of a symmetric
traceless O(N) tensor field $A_2$ by
\eqn\eatwo{A_2 \equiv Z_{A_2} \left( \Phi_0^N \Phi_0^N - {1 \over N} \right) .}
The O(N) symmetry restricts the RG equations in the following form:
\eqn\eRGmatrix{\dt \Phi = \Gamma (g,m^2) \Phi ,}
where
\eqnn\ePhihere
\eqnn\egamma
$$\eqalignno{&\Phi \equiv (\one, A_1,  A_2, B_0, \p^2 A_1)^T ,&\ePhihere\cr
&\Gamma (g,m^2)
\equiv \pmatrix{0&0&0&0&0\cr 0& \gamma_{A_1}&0&0&0\cr
0&0&\gamma_{A_2}&0&0\cr
0& m^2 \gamma_{B_0}^{~A_1}& 0&2+ \gamma_{B_0}&0\cr
0&0&0&0&2 + \gamma_{A_1}\cr} .&\egamma\cr}$$
Note that Eqs.~\ezrel, \econj, and \eRGconj\ imply
\eqn\egammarel{\eqalign{
\gamma_{A_1} &= {\beta_g \over g} - \beta_m \cr
\gamma_{B_0}^{~A_1} &= g {d \over dg} \left(
- \beta_m + {\beta_g \over g} \right) \cr
\gamma_{B_0} &= - g^2 {d \over dg} {\beta_g \over g^2}~ .\cr}}
Similarly, the O(N) invariance implies
\eqn\eopem{(C_m) (r;g) =
\pmatrix{0&0&0&0&0\cr 0&0&0&0&0\cr 0&0&0&0&0\cr
0&(C_m)_{B_0}^{~A_1} (r;g)& 0&0&0\cr
(C_m)_{\p^2 A_1}^{~\one} (r;g)& 0&
(C_m)_{\p^2 A_1}^{~A_2} (r;g)&0&0\cr} ,}
and the nonvanishing elements of the matrix $C_g (r;g,m^2)$ are
\eqn\eopeg{\eqalign{&(C_g)_{A_1}^{~A_1} (r;g) ,\quad
(C_g)_{A_2}^{~A_2} (r;g) ,\cr
& (C_g)_{B_0}^{~\one} (r;g) ,\quad (C_g)_{B_0}^{~A_1} (r;g,m^2) ,
\quad (C_g)_{B_0}^{~B_0} (r;g) ,\cr
& (C_g)_{\p^2 A_1}^{~\one} (r;g,m^2) ,\quad
(C_g)_{\p^2 A_1}^{~A_1} (r;g) ,\quad
(C_g)_{\p^2 A_1}^{~A_2} (r;g,m^2) ,\quad
(C_g)_{\p^2 A_1}^{~\p^2 A_1} (r;g) ,\cr}}
where we have taken the angular average over $r$, and the terms
less singular than $1/r^2$ have been dropped.  All the unwritten elements
of $C_g$ vanish.  Following \eh\ we define
\eqn\ehdef{H_m (g) \equiv C_m (r=1;g) ,\quad
H_g (g,m^2) \equiv C_g (r=1;g,m^2) .}
We introduce the connection ($c_m (g), c_g (g,m^2)$), which has non-vanishing
elements exactly where
the OPE coefficients have non-vanishing elements.

In this particular model, Eq.~\eformula\ implies
\eqn\eformulahere{\eqalign{
H_m (g) &= {\p \over \p m^2} \Psi + [c_m, \Psi] + \beta_g \Omega_{gm} ,\cr
H_g (g,m^2) &= {\p \over \p g} \Psi + [c_g, \Psi] - (2 + \beta_m) m^2
\Omega_{gm} ,\cr}}
where
\eqnn\epsihere
\eqnn\eomegahere
$$\eqalignno{\Psi (g,m^2) &\equiv
\Gamma + (2 + \beta_m) m^2 c_m + \beta_g c_g ~,&\epsihere\cr
\Omega_{gm} (g) &\equiv
{\p \over \p g} c_m - {\p \over \p m^2} c_g + [c_g, c_m] ~.&\eomegahere\cr}$$
Eq.~\ecurvformula\ gives
\eqn\ecurvintegral{\eqalign{&\Omega_{gm} (g) \vev{\Phi}_{g,m^2} =
\int_{1 \ge r} d^2 r ~{\rm F.P.} \int_{1 \ge r'} d^2 r' \cr
&\quad \times \vev{ \left( \O_m (r) \left( \O_g (r') - {1 \over 2 \pi} C_g (r')
\right)
- (r \leftrightarrow r') \right) \Phi (0) }_{g,m^2}^c .\cr}}

In the next two sections we will compute the OPE coefficients
$H_m, H_g$ and the curvature $\Omega_{gm}$ perturbatively
and determine the connection $c_m, c_g$ by solving
Eqs.~\eformulahere\ and \eomegahere.

\newsec{Perturbative calculations}

The technique of perturbative calculations is well-known.
In order to obtain OPE coefficients in the coordinate
space \ref\rcollins{See, for example, J.~Collins,
{\it Renormalization} (Cambridge University Press, 1984).}, we have used the
formula
\eqn\edimreg{\int {d^{2+\ep} k \over (2\pi)^{2+\ep}}~\mu^{-\ep}
{{\rm e}^{i k r} \over (k^2)^n} = {1 \over \pi r^2}
\left( {r \over 2} \right)^{2n}
\left({{\rm e}^{\gamma\over 2} \over r} \right)^{\ep}
{\Gamma \left(1+{\ep \over 2} - n\right)
\over \Gamma (n)} ,}
where we have chosen the renormalization scale as
\eqn\emu{\mu^2 = {{\rm e}^{- \gamma} \over \pi} }
so that
\eqn\emutwo{\int {d^{2+\ep} k \over (2\pi)^{2+\ep}}~
\mu^{-\ep} ~{{\rm e}^{i k r} \over k^2 } - {1 \over 2 \pi \ep} = - {1 \over 2
\pi}~\ln r}
in the limit $\ep \to 0$.

We only give the final results, leaving the intermediate results
to appendices A,B.
First, we find the following beta functions and anomalous dimensions:
\eqn\ebetagamma{\eqalign{
\beta_m &\simeq {g \over \pi} {N-3 \over 4} +
\left({g \over \pi}\right)^2 {N-2\over 4} \cr
\beta_g &\simeq {g^2 \over \pi} {N-2\over 2} + {g^3\over \pi^2} {N-2\over 4}\cr
\gamma_{A_1} &= {g \over \pi} {N-1\over 4} + {\rm O} (g^3) ~,\quad
\gamma_{A_2} = {g \over \pi} {N\over 2} + {\rm O} (g^3) \cr
\gamma_{B_0} &\simeq {g^2 \over \pi^2} {-N+2\over 4} ~,\quad
\gamma_{B_0}^{~A_1} = {g \over \pi} {N-1\over 4} + {\rm O} (g^3) .\cr}}
Second, we find the following results for $H_m, H_g$:
\eqnn\ehmb
\eqna\ehmda
$$\eqalignno{(H_m)_{B_0}^{~A_1} (g) &= {g\over \pi} {N-1\over 4}
( 1 + {\rm O} (g^2)) &\ehmb\cr
(H_m)_{\p^2 A_1}^{~\one} &\simeq {g \over \pi} {-(N-1)\over 2 N\pi} \left(
1 + {g \over \pi} (N-1) a \right)&\ehmda a\cr
(H_m)_{\p^2 A_1}^{~A_2} &\simeq {g\over \pi} {-(N-1)\over 2} \left(
1 - {g \over \pi} a \right) ~,&\ehmda b\cr}$$
where $a$ is an unknown constant, and
\eqnn\ehgaone
\eqnn\ehgatwo
\eqna\ehgb
\eqna\ehgda
$$\eqalignno{
(H_g)_{A_1}^{~A_1} (g) &= {N-1 \over 4 \pi} (1 + {\rm O}(g^2) )&\ehgaone\cr
(H_g)_{A_2}^{~A_2} (g) &= {N \over 2 \pi} (1 + {\rm O}(g^2)) &\ehgatwo\cr
(H_g)_{B_0}^{~B_0} (g) &\simeq {-(N-2) \over 2 \pi} &\ehgb a\cr
(H_g)_{B_0}^{~\one} (g) &\simeq - {(N-1) \over 2 \pi} &\ehgb b\cr
(H_g)_{B_0}^{~A_1} (g,m^2) &\simeq m^2 ~{g
\over \pi}~ {-(N-1)(N-2)\over 8 \pi} &\ehgb c\cr
(H_g)_{\p^2 A_1}^{~\p^2 A_1} (g) &\simeq {-(N-3)\over 4\pi}&\ehgda a\cr
(H_g)_{\p^2 A_1}^{~A_1} (g) &\simeq {N-1\over \pi}
\left( 1 - {g \over \pi} (N-2) \right)&\ehgda b\cr
(H_g)_{\p^2 A_1}^{~\one} (g,m^2) &\simeq m^2~ {-(N-1)(N-3)\over
4 \pi N} \left( 1 + {g \over \pi} ~b_1 \right)&\ehgda c\cr
(H_g)_{\p^2 A_1}^{~A_2} &\simeq m^2~ {3(N-1)\over 4 \pi} \left(
1 + {g \over \pi} ~b_2 \right) ~,&\ehgda d\cr}$$
where the constants $b_1, b_2$ are related by
\eqn\ebrel{ - (N-3) b_1 + 3 (N-1) b_2 = N(N-2)~.}
The unknown constants $a, b_1, b_2$ appear due to
\eqn\ereloneatwo{A_2 - {N-1\over N} ~\one = {\rm O}(g) ~.}
In the next section we will be able to relate $b_1, b_2$ to the constant $a$.
Third, from Eq.~\ecurvintegral\ we find the following curvature
(see appendix C for more details):
\eqn\ecurvpert{\eqalign{&\quad(\Omega_{gm})_{B_0}^{~A_1} (g)
\simeq {N-1 \over 8 \pi} \cr
&(\Omega_{gm})_{\p^2 A_1}^{~\one} (g) +
(\Omega_{gm})_{\p^2 A_1}^{~A_2} (g) {N-1 \over N}
\simeq {-(N-1) \over 4 \pi} ~.\cr}}

\newsec{Determination of the connection}

The matrix elements of the connection $c_g, c_m$
depend on the convention adopted.  First of all,
under a change of basis
\eqn\echangebase{\Phi \to \Phi' \equiv N(g,m^2) \Phi~,}
the connection transforms as
\eqn\echangeconni{\eqalign{c_g &\to
c_g' = N(g,m^2) \left( \p_g N(g,m^2)^{-1} + c_g (g,m^2) N(g,m^2)^{-1}
\right) \cr
c_m &\to c_m' = N(g,m^2)
\left( \p_{m^2} N(g,m^2)^{-1} + c_m (g) N(g,m^2)^{-1} \right) ~.\cr}}
Second, under a coordinate change
\eqn\echange{g \to \tilde{g} \equiv f(g)~,\quad
m^2 \to \tilde{m}^2 \equiv m^2 h(g)~,}
where
\eqn\efh{f'(0) = h(0) = 1~,}
the conjugate fields transform as
\eqna\echangeconj
$$\eqalignno{\O_g \to \O_{\tilde{g}} &=
\left({\p g \over \p \tilde{g}} \right)_{\tilde{m}^2} \O_g +
\left({\p m^2 \over \p \tilde{g}} \right)_{\tilde{m}^2} \O_m\cr
&= {1 \over f'(g)} \left( \O_g - {h'(g) \over h(g)}~ m^2 \O_m \right)
&\echangeconj a\cr
\O_m \to \O_{\tilde{m}} &= \left( {\p g \over \p \tilde{m}^2}
\right)_{\tilde{g}}
\O_g + \left({\p m^2 \over \p \tilde{m}^2} \right)_{\tilde{g}} \O_m
= {1 \over h(g)} \O_m ~.&\echangeconj b\cr}$$
If we do not change the basis $\Phi$, then the connection
simply transforms as
\eqn\echangeconnii{c_{\tilde{g}} = {1 \over f'(g)} \left(
c_g - {h'(g)\over h(g)} m^2 c_m \right)~,\quad
c_{\tilde{m}} = {1 \over h(g)}~c_m~.}
Since the conjugate fields transform as \echangeconj{},
the connection for the conjugate fields
transforms inhomogeneously as
\eqn\echangeconniii{\eqalign{
(c_{\tilde{g}})_{\tilde{g}}^{~\tilde{g}} &= {1 \over f'(g)} \left( (c_g)_g^{~g}
+ {f''(g) \over f'(g)}
\right)\cr
(c_{\tilde{g}})_{\tilde{g}}^{~\tilde{m}} &= {\tilde{m}^2 \over (f'(g))^2}
\Bigg(
{h'(g) \over h(g)} \left( (c_g)_g^{~g} - 2 (c_g)_m^{~m} \right)
+ {1 \over m^2} (c_g)_g^{~m} \cr
&\qquad\qquad\qquad\qquad\qquad + {h''(g) \over h(g)} -
2 \left( {h'(g)\over h(g)} \right)^2 \Bigg) \cr
(c_{\tilde{g}})_{\tilde{g}}^{~\one} &= {1 \over f'(g)^2} (c_g)_g^{~\one} \cr
(c_{\tilde{g}})_{\tilde{m}}^{~\tilde{m}} &=
(c_{\tilde{m}})_{\tilde{g}}^{~\tilde{m}} = {1 \over f'(g)}
\left(  (c_m)_g^{~m} + {h'(g) \over h(g)} \right) ~.\cr}}
The results we present below correspond to
a specific choice of the parameters
$g, m^2$ and fields $\Phi$ in the MS scheme with
the renormalization scale $\mu$ chosen by Eq.~\emu.

Using the OPE coefficients $H_m, H_g$ and
the curvature $\Omega_{gm}$ that we have computed
in the previous section, we can determine
the connection $c_m, c_g$ from Eqs.~\eformulahere,
\eomegahere.

We find the following matrix elements for $c_m$:
\eqnn\eccmb
\eqna\eccmda
$$\eqalignno{(c_m)_{B_0}^{~A_1} (g) &= {\rm O}(g^2) &\eccmb\cr
(c_m)_{\p^2 A_1}^{~\one} (g) &\simeq {N-1\over N}
\left( 1 - {g \over \pi}
{1\over 2(N-3)} ~K \right) &\eccmda a\cr
(c_m)_{\p^2 A_1}^{~A_2} (g) &\simeq -1 + {g \over \pi} {1 \over 2(2N-3)} ~K~,
&\eccmda b\cr}$$
where the constant $K$ is related to the constant $a$
in Eqs.~\ehmda{a,b}\ by
\eqn\eK{K \equiv N-2 + 2 a (N-1)~.}
We cannot determine the constant $K$ to the order
we have calculated.

We find the following matrix elements for $c_g$:
\eqnn\eccga
\eqna\eccgb
\eqna\eccgda
$$\eqalignno{
(c_g)_{A_1}^{~A_1} (g) &= {\rm O} (g)~,\quad
(c_g)_{A_2}^{~A_2} (g) = {\rm O} (g) &\eccga\cr
(c_g)_{B_0}^{~B_0} (g) &\simeq - {1 \over g} + {N-2\over 4 \pi}&\eccgb a\cr
(c_g)_{B_0}^{~\one} (g) &\simeq {N-1\over 4\pi} \left( 1 - {g \over \pi}
{N-2\over 2}
\right) &\eccgb b\cr
(c_g)_{B_0}^{~A_1} (g,m^2) &\simeq - m^2~{N-1 \over 8 \pi} &\eccgb c\cr
(c_g)_{\p^2 A_1}^{~\p^2 A_1} (g) &\simeq - {1 \over g} &\eccgda a\cr
(c_g)_{\p^2 A_1}^{~A_1} (g) &\simeq - {N-1\over 2 \pi}
\left( 1 - {g \over \pi} {N-2\over 2} \right) &\eccgda b\cr
(c_g)_{\p^2 A_1}^{~\one} (g,m^2) &\simeq - {m^2 \over g} {N-1\over N}
\left( 1 - {g \over \pi}  {N-1-{K \over 2N-3}\over 4} \right) &\eccgda c\cr
(c_g)_{\p^2 A_1}^{~A_2} (g,m^2) &\simeq {m^2 \over g}
\left( 1 + {g \over \pi} {1 + {K \over 2N-3} \over 4} \right)~, &\eccgda
d\cr}$$
where we could eliminate the constants $b_1, b_2$,
which are now related to the sole constant $K$ of Eq.~\eK\ by
\eqn\ebonebtwoK{\eqalign{b_1 &= {1 \over N-3}
\left( -N^2 + {11 \over 2}~N - {13 \over 2} + {2(N-2)\over 2N-3}~K
\right) \cr
b_2 &= {1 \over 3(N-1)}
\left( {7N-13\over 2} + {2(N-2)\over 2N-3}~K\right) ~.\cr}}
thanks to Eq.~\eomegahere.

Finally, using Eqs.~\econj\ and the transformation
properties \echangeconni, we obtain
\eqna\ecgcmpert
$$\eqalignno{&(c_g)_m^{~m} (g) = (c_m)_g^{~m} (g) =
{1 \over g} + (c_g)_{A_1}^{~A_1} (g)
= {1 \over g} + {\rm O}(g) &\ecgcmpert a\cr
&(c_g)_g^{~g} (g) = {2 \over g} + (c_g)_{B_0}^{~B_0} (g)
\simeq {1 \over g} + {N-2\over 4 \pi}&\ecgcmpert b\cr
&(c_g)_g^{~\one} (g) = - {1 \over g^2} (c_g)_{B_0}^{~\one} (g)
\simeq - {1 \over g^2}~{N-1\over 4 \pi} + {1 \over g}~
{(N-1)(N-2)\over 8 \pi^2}&\ecgcmpert c\cr
&(c_g)_g^{~m} (g,m^2) = {1 \over g} \left(
(c_g)_{B_0}^{~A_1} (g,m^2)  + m^2 \left( (c_g)_{B_0}^{~B_0} (g)
- (c_g)_{A_1}^{~A_1} (g) \right) \right) \cr
&\qquad\qquad \simeq m^2 \left(
- {1 \over g^2} + {1 \over g}~{N-3\over 8\pi} \right) ~.  &\ecgcmpert d\cr}$$

\newsec{Zero temperature limit}

By examining the connection $c_m, c_g$ obtained
in the previous section, we notice that many matrix
elements diverge at zero temperature $g=0$.
This is disturbing; the connection controls the
short-distance properties of the theory, and we expect
their matrix elements to behave in the same way as
the beta functions and anomalous dimensions
at low temperatures, i.e., we expect them to be
finite polynomials of the mass parameter $m^2$
whose coefficients are smooth functions of $g$
and admit perturbative expansions around $g=0$.
But most of the divergences at $g=0$ are simply
due to the wrong normalization of the composite fields.

We can see this as follows.  Consider composite
fields $\Phi_a$.  Suppose the diagonal elements
of the connection $c_g$ diverge as
\eqn\ediv{(c_g)_{\Phi_a}^{~\Phi_b} (g) =  {n \over g} ~\delta_a^{~b}+ {\rm
O}\left(g^0\right)~.}
Then Eq.~\echangeconni\ implies that
for the renormalized fields $g^n \Phi^a$
the diagonal elements become finite:
\eqn\efinite{(c_g)_{g^n \Phi_a}^{~g^n \Phi_b} (g) =
- {n \over g}~\delta_a^{~b} + (c_g)_{\Phi_a}^{~\Phi_b} (g) =
{\rm O} \left( g^0\right)~.}

We can identify another source of divergences in the connection
$c_g, c_m$ as the wrong normalization of the parameter
$m^2$.  The field $\O_m$, which is conjugate to $m^2$, is given by
the second of Eqs.~\econj.  The finiteness of $(c_g)_{A_1}^{~A_1} (g)$
at $g=0$ implies that it is the field $A_1$, but not $\O_m$,
which has proper normalization.  Hence, we should redefine the mass parameter
(or external magnetic field) by
\eqn\emass{\tilde{m}^2 \equiv {m^2 \over g} ~.}
This is nothing but a reduced external magnetic field.
Then, Eq.~\echangeconj{b}\ gives the new
conjugate field as
\eqn\enewom{\O_{\tilde{m}} = g~\O_m = - A_1 ~,}
where we have kept the temperature intact:
\eqn\eg{\tilde{g} = g~.}
Eq.~\echangeconj{a}\ gives
\eqn\enewog{\O_{\tilde{g}} = \O_g + {m^2\over g}~\O_m = - {1 \over g^2}~B_0~.}

We now redefine the fields $\Phi^T = (\one, A_1, A_2, B_0, \p^2 A_1)$
by
\eqn\enewfield{\Phi'^T = \left(\one, A_1, A_2,
B'_0 \equiv {1 \over g} B_0, {1 \over g} \p^2 A_1
\right)~,}
and use $\tilde{g} \equiv g$ and $\tilde{m}^2$ as the parameters.
Then we find the following
matrix elements for the connection $c_{\tilde{m}}, c_{\tilde{g}}$:
\eqnn\enewcmb
\eqna\enewcmda
\eqnn\enewcgaone
\eqnn\enewcgatwo
\eqna\enewcgb
\eqna\enewcgda
$$\eqalignno{&(c_{\m})_{B'_0}^{~A_1} (g) = {\rm O} (g^2) &\enewcmb\cr
&(c_{\m})_{{1 \over g} \p^2 A_1}^{~\one} (g) \simeq {N-1\over N} &\enewcmda
a\cr
&(c_{\m})_ {{1 \over g} \p^2 A_1}^{~A_2} (g) \simeq - 1&\enewcmda b\cr
&(c_{\g})_{A_1}^{~A_1} (g) = {\rm O} (g) &\enewcgaone\cr
&(c_{\g})_{A_2}^{~A_2} (g) = {\rm O} (g) &\enewcgatwo\cr
&(c_{\g})_{B'_0}^{~B'_0} (g) \simeq
{N-2 \over 4 \pi} &\enewcgb a\cr
&(c_{\g})_{B'_0}^{~\one} (g) \simeq {1 \over g} {N-1 \over 4 \pi}
\left( 1 - {g \over \pi} {N-2\over 2} \right) &\enewcgb b\cr
&(c_{\g})_{B'_0}^{~A_1} (g,\m^2) \simeq - {N-1 \over 8 \pi} \m^2 &\enewcgb c\cr
&(c_{\g})_{{1 \over g} \p^2 A_1}^{~{1 \over g}\p^2 A_1} (g) = {\rm O} (1)
&\enewcgda a\cr
&(c_{\g})_{{1 \over g} \p^2 A_1}^{~A_1} (g) \simeq {1 \over g} {-(N-1)
\over 2 \pi} &\enewcgda b\cr
&(c_{\g})_{{1 \over g} \p^2 A_1}^{~\one} (g,\m^2)
= {\m}^2 {\rm O} (g^0) &\enewcgda c\cr
&(c_{\g})_{{1 \over g} \p^2 A_1}^{~A_2} (g,\m^2)
= {\m}^2 {\rm O} (g^0) ~.&\enewcgda d\cr}$$
We find that all the matrix elements,
except the $(B'_0, \one)$ and $({1 \over g}\p^2 A_1, A_1)$
elements of $c_{\g}$, become finite
at $g=0$ as functions of $g, {\m}^2$.
All the maximal elements, i.e., those relating
two fields which can mix under the RG, are
finite at $g=0$.  In fact the finiteness of
the maximal elements at $g=0$ specifies
the new basis $\Phi'$ uniquely up to
a linear transformation \echangebase,
where $N$, as a function of $g, \tilde{m}^2$,
must be invertible even at the origin $g=0$.
The proper normalization of the basis is necessary
in order to assure that the vector bundle of
the composite fields is well-defined in a neighborhood
of the origin $g=0$.  We can also explain the singularities
of the non-maximal elements \enewcgb{b}\ and \enewcgda{b},
but we leave it to appendix D since
the explanation requires the familiarity with the
variational formula \rsonodaiii, which we have not
introduced in this paper.

We also find, from Eqs.~\echangeconniii, the following matrix elements
of the connection for the redefined conjugate fields:
\eqna\enewcgcm
$$\eqalignno{&(c_{\g})_{\g}^{~\g} (g) = (c_g)_g^{~g} (g)
\simeq {1 \over g} + {N-2\over 4\pi} &\enewcgcm a\cr
&(c_{\g})_{\g}^{~\m} (g,{\m}^2) = {\m}^2 \left(
- {1 \over g} \left( (c_g)_g^{~g} - 2 (c_g)_m^{~m} \right)
+ {1 \over m^2} (c_g)_g^{~m} \right) \cr
&\qquad\qquad\qquad \simeq {{\m}^2 \over g} {-N+1 \over 8 \pi} &\enewcgcm b\cr
&(c_{\g})_{\g}^{~\one} (g) = (c_g)_g^{\one} (g) \simeq - {1 \over g^2}
{N-1 \over 4 \pi} \left( 1 - {g \over \pi} {N-2\over 2} \right) &\enewcgcm c\cr
&(c_{\g})_{\m}^{~\m} (g) = (c_{\m})_{\g}^{~\m} (g) = (c_m)_g^{~m} - {1 \over g}
= {\rm O} (g)~. &\enewcgcm d\cr}$$
We have two maximal elements to pay attention to:
$(c_{\g})_{\g}^{~\g}$ and $(c_{\g})_{\m}^{~\m}$.
Only the second is finite at zero temperature.  The divergence
of the element $(c_{\g})_{\g}^{~\g}$ comes from the singular
relation of the conjugate field $\O_{\g}$ to the
well-normalized field $B'_0$:
\eqn\eogbzero{
\O_{\g} = - {1 \over g}~B'_0 ~.}
The redefinition of $\g = g$ by
\eqn\egprime{g' \equiv \ln g}
makes the diagonal element $(c_{g'})_{g'}^{~g'}$ regular at $g=0$,
but the non-analytic transformation \egprime\ is illegal
as a coordinate transformation near the origin $g=0$.
At the end of appendix D we give
a technical reason for the necessity of ${1\over g}$ in Eq.~\eogbzero:
without the factor the conjugate field is a total derivative
at $g=0$, and it does not deform the theory.
We reconcile ourselves with the zero-temperature
singularity of $(c_{\g})_{\g}^{~\g}$ by regarding
it as a unique feature of a theory with its
symmetry non-linearly realized: we can trace the source
of the singularity to the bare temperature $g_0$ in the
denominator of the lagrangian \eL.

Finally, we must understand the physical meaning of
the normalization of the redefined composite fields
$\Phi'$ (Eq.~\enewfield).  We will show that the fields
$\Phi'$ have well-defined and non-trivial correlation functions
at low temperatures in the {\bf absence}
of an external magnetic field, i.e., $m^2 = 0$.
We do not have a general proof, and
we will be content with verifying this claim
only for the two-point functions.
Perturbative calculations give the following results:
\eqnn\etwoaone
\eqnn\etwob
\eqnn\etwoda
$$\eqalignno{\vev{\Phi^I (r) \Phi^J (0)}_g &= {\delta^{IJ} \over N}
\left( 1 + {g \over \pi} {1-N\over 2}~\ln r + {\rm O}(g^2) \right)
&\etwoaone\cr
\vev{B'_0 (r) B'_0 (0)}_g &= {N-1 \over 4 \pi^2 r^4}
+ {\rm O} (g) &\etwob\cr
\vev{{1 \over g} \p^2 \Phi^I (r) {1 \over g} \p^2 \Phi^J (0)}_g \cr
= \delta^{IJ}
{1 \over \pi^2 r^4}{N-1\over N}&
\left( 1 + {g \over \pi} (N-3) \left( {1 \over 2} \ln r - 1\right)
+ {\rm O} (g^2) \right) ~.
&\etwoda\cr}$$
(We did not try to compute the two-point function of
$A_2$.)  We recall that the correlation functions must be
O(N) invariant at $m^2 = 0$.

Hence, we find that the connection $c_{\g}, c_{\m}$ has
finite matrix elements (excluding the non-maximal elements)
at $g=0$ if we normalize the fields so that their correlation
functions have well-defined zero-temperature limits.

In order to make the above finding more firm, let us consider
one more field that we have not considered before, i.e.,
the spatial derivative of the O(N) vector, $\p_\mu \Phi^I$.
By taking the second-order derivative of \etwoaone, we obtain
\eqn\etwodaone{
\vev{{1 \over \sqrt{g}} \p_\mu \Phi^I (r)
{1 \over \sqrt{g}} \p_\nu \Phi^J (0)}_g \simeq \delta^{IJ}
{1 \over \pi r^2} {N-1 \over 2 N} \left(
\delta_{\mu\nu} - 2 {r_\mu r_\nu \over r^2} \right)~.}
Therefore, the field ${1 \over \sqrt{g}} \p_\mu \Phi^I$
is properly normalized.  To show that
the connection $c_{\g}$ for ${1 \over \sqrt{g}} \p_\mu \Phi$
is finite at $g=0$, we must compute the OPE coefficient in
\eqn\ecgdaone{
\O_{\g} (r) \p_\mu \Phi^I (0) \simeq {1 \over 2 \pi}
(C_{\g})_{\p_\mu \Phi^I}^{~\p_\nu \Phi^J} (r;g) \p_\nu \Phi^J (0)~.}
This is done in appendix E.  We find
\eqn\ehgdaone{\eqalign{
&(H_{\g})_{{1 \over \sqrt{g}} \p_\mu \Phi^I}^{~{1 \over \sqrt{g}} \p_\nu
\Phi^J} (g)
= (H_g)_{\p_\mu \Phi^I}^{~\p_\nu \Phi^J} (g) \cr
&\quad = \delta^I_{~J} \delta_{\mu}^{~\nu} {1 \over 2}
\left( (H_g)_{A_1}^{~A_1} (g) + (H_g)_{\p^2 A_1}^{~\p^2 A_1} (g) \right)
\simeq \delta^I_{~J} \delta_{\mu}^{~\nu} {1 \over 8\pi}~.\cr}}
Since
\eqn\egammadphi{{d \over dt} {1 \over \sqrt{g}} \p_\mu \Phi^I =
\delta^I_{~J} \delta_\mu^{~\nu}
\left(1 + \gamma_{A_1} - {\beta_g \over 2 g} \right) {1 \over \sqrt{g}} \p_\nu
\Phi^J~,}
Eqs. \eformulahere\ imply
\eqn\ehgdaonerel{
(H_{\g})_{{1 \over \sqrt{g}} \p_\mu \Phi^I}^{~{1 \over \sqrt{g}} \p_\nu \Phi^J}
(g)
= {d \over dg} \left(  \delta^I_{~J} \delta_\mu^{~\nu}
\left(\gamma_{A_1} - {\beta_g \over 2 g}\right)
+ \beta_g~(c_{\g})_{{1 \over \sqrt{g}} \p_\mu \Phi^I}^{~{1 \over \sqrt{g}}
\p_\nu \Phi^J}
\right) ~.}
Hence, we find, from Eqs.~\ebetagamma\ and \ehgdaone,
\eqn\ecgdaone{(c_{\g})_{{1 \over \sqrt{g}} \p_\mu \Phi^I}^{~{1 \over \sqrt{g}}
\p_\nu \Phi^J}
= \delta^I_{~J} \delta_\mu^{~\nu}~{\rm O} (g^0)~.}
Thus, the connection is finite at $g=0$ for the field
${1 \over \sqrt{g}} \p_\mu \Phi^I$, which has a well-defined
two-point function in the limit $g=0$.

\newsec{Conclusion}

In this paper we have calculated the connection $c_g, c_m$
for the composite fields $\Phi$ (Eq.~\ePhihere).  From the behavior
of the matrix elements near zero temperature $g=0$, we have
concluded that the reduced magnetic field $\tilde{m}^2 \equiv
{m^2 \over g}$ is more natural at low temperatures, and that
the matrix elements are finite at $g=0$ only if we normalize
the composite fields so that their correlation
functions have well-defined and
non-trivial zero temperature limits.
Since we take these limits in the absence of an
external magnetic field,
the correlation functions are O(N) invariant even at zero
temperature.

We find especially the second conclusion pleasing, since
the connection controls the short-distance
physics (equivalently low temperature physics) together
with the beta functions and anomalous dimensions.
Anything to do with short-distance physics should be
given as finite polynomials of the reduced magnetic field $\tilde{m}^2$
whose coefficients can be formally expanded in powers of $g$.

We believe it would be
interesting to examine non-abelian gauge theories which
also realize the relevant symmetry
non-linearly \ref\rsuzarkesh{
W.-C.~Su and A.~Zarkesh, work in progress}.
We expect to find that the behavior of the connection at
short distances dictates how to take the short-distance limit
of the BRST invariant theories.

\vskip 0.2in
H. S. thanks Terry Tomboulis for discussions.

\appendix{A}{Renormalized composite fields}

For perturbative calculations, we introduce $N - 1$ independent
scalar fields $\phi^i (i=1,...,N-1)$ such that
\eqn\ephifield{\Phi^i = \sqrt{g}~\phi^i \quad
(i=1, ... ,N-1) ,\quad \Phi^N = \sqrt{1 - Z_\Phi g \vec{\phi}^2
\over Z_\Phi} ~.}

The following composite fields are renormalized:
\eqnn\ephione
\eqnn\ephitwo
\eqnn\edphi
\eqnn\eddphi
$$\eqalignno{
&\bra{\phione}_1 \equiv \left( 1 - {g \over 2 \pi \ep} \right) \left( \phione
+ {N-1 \over 4 \pi \ep} \right) &\ephione\cr
&\bra{\phitwo}_0 \equiv \phitwo + {N+1 \over 4 \pi \ep}
\phione + {(N-1)(N+1) \over 32 \pi^2 \ep^2} &\ephitwo\cr
&\bra{\dphi}_1 \equiv \dphi - {N-1 \over 4 \pi \ep} m^2 \cr
&\quad + g \left( {N-2 \over 2 \pi \ep} \dphi + m^2 {N+1 \over 4 \pi \ep}
\phione
- m^2 {(N-1)(N-5) \over 16 \pi^2 \ep^2} \right)&\edphi\cr
&\bra{\ddphi}_0 \equiv \ddphi + {1 \over 2 \pi \ep} \dphi \cr
&\quad\quad - {m^2 \over 2 \pi \ep} \phione -
{(N-1)m^2 \over 8 \pi^2 \ep^2} ~,&\eddphi\cr}$$
where $[({\rm field})]_n$ is renormalized to
order $g^n$.
Since
\eqn\ederivative{\p^2 \phione = 2 \left( \dphi + m^2 \phione \right) + {\rm O}
(g) ,}
we find
\eqn\edphitwo{\p^2 \bra{\phitwo}_0 \equiv \p^2 \phitwo + {N+1 \over 2 \pi \ep}
\left( \dphi + m^2 \phione \right) ~.}

Using Eqs.~\edimreg, \emu, and \emutwo, we can compute the OPE
of the above renormalized composite fields as follows:
\eqna\eope
$$\eqalignno{&\bra{\dphi}_1 (r) \bra{\phione}_1 (0) \cr
&\quad\quad \simeq
{N-1 \over 8 \pi^2 r^2} \left( 1 + {g \over \pi} (N-2) \ln r \right) \one
 - g {N+1 \over 4 \pi^2 r^2} \bra{\phione}_0 (0) &\eope a\cr
&\bra{\dphi}_0 (r) \bra{\phitwo}_0 (0) \simeq {N+1\over 8 \pi^2 r^2}
\bra{\phione}_0 (0) &\eope b\cr
&\bra{\ddphi}_0 (r) \bra{\phione}_0 (0) \simeq {1 \over 4 \pi^2 r^2}
\bra{\phione}_0 (0)
&\eope c\cr
&\bra{\dphi}_1 (r) \bra{\dphi}_1 (0) \simeq (N-1) \Bigg[
\left( {1 \over 4 \pi^2 r^4} - {m^2 \over 8 \pi^2 r^2} \right) \cr
&\qquad -  {g \over 4 \pi^3} \left( {N-2 \over r^4} (1 - \ln r)
+ {m^2 \over 4 r^2} (2-N+(N-3) \ln r) \right) \Bigg] \one &\eope d\cr
&\quad + {g \over \pi^2 r^2} \left(
\left({1 \over r^4} + {m^2 \over r^2} {3N+1 \over 8} \right) \bra{\phione}_0
(0)
+ {2N-5\over 4} \bra{\dphi}_0 (0) \right) \cr
&\bra{\dphi}_0 (r) \bra{\ddphi}_0 (0) \cr
&\quad \simeq \left( {1 \over 2 \pi^2 r^4} - {m^2 \over 4 \pi^2 r^2} \right)
\bra{\phione}_0 (0)
+ {1 \over 4 \pi^2 r^2} \bra{\dphi}_0 (0) &\eope e\cr
&\bra{\ddphi}_0 (r) \bra{\dphi}_0 (0) \cr
&\quad \simeq \left( {1 \over 2 \pi^2 r^4}
- {m^2 N \over 4 \pi^2 r^2} \right) \bra{\phione}_0 (0)
+ {-N+2 \over 4 \pi^2 r^2} \bra{\dphi}_0 (0) &\eope f\cr
&\p^2 \bra{\phitwo}_0 (r) \bra{\phione}_0 (0)
\simeq {N+1 \over 4 \pi^2 r^2} \bra{\phione}_0 (0)&\eope g\cr
&\bra{\dphi}_0 (r) \p^2 \bra{\phitwo}_0 (0) \cr
&\qquad \simeq {N+1 \over \pi^2} \left(
\left( {1 \over 2 r^4} + {m^2 \over 4 r^2} \right) \bra{\phione}_0 (0)
+ {1 \over 4 r^2} \bra{\dphi}_0 (0) \right) &\eope h\cr
&\p^2 \bra{\phitwo}_0 (r) \bra{\dphi}_0 (0) \simeq
{N+1 \over 2 \pi^2 r^4} \bra{\phione}_0 (0) ~, &\eope i\cr}$$
where we have taken the angular average over $r$.

\appendix{B}{OPE coefficients $C_m, C_g$}

The O(N) covariant composite fields are related to the composite fields
of appendix A in the following way:
\eqnn\eaone
\eqnn\eatwo
\eqnn\ebzero
\eqnn\edaone
\eqnn\eom
\eqnn\eog
$$\eqalignno{A_1 &\simeq 1 - g \bra{\phione}_1 - g^2 \bra{\phitwo}_0 &\eaone\cr
A_2 &\simeq {N-1\over N} - 2 g \bra{\phione}_1 &\eatwo\cr
B_0 &\simeq g \left( \bra{\dphi}_1 + g \bra{\ddphi}_0 \right)&\ebzero\cr
\p^2 A_1 &\simeq g \left( - 2 \bra{\dphi}_1 - 2 m^2 \bra{\phione}_1 \right) \cr
&\quad + g^2 \left( \p^2 \bra{\phitwo}_0 - 4 \bra{\ddphi}_0
- 4 m^2 \bra{\phitwo}_0 \right) &\edaone\cr
\O_m &\simeq - {1 \over g} + \bra{\phione}_1 + g \bra{\phitwo}_0 &\eom\cr
\O_g &\simeq - {1 \over g} \bra{\dphi}_1 - \bra{\ddphi}_0 \cr
&\qquad\qquad + {m^2 \over g}
- {m^2 \over g} \bra{\phione}_1 - m^2 \bra{\phitwo}_0 ~.&\eog\cr}$$

Now it is straightforward to
compute the coefficient functions in the products
$\O_m (r) \Phi (0)$ and $\O_g (r) \Phi (0)$ using
the above relations and Eqs.~\eope{}.
The OPE coefficients $C_m$ are obtained as
\eqnn\ecmb
\eqna\ecmda
$$\eqalignno{
(C_m)_{B_0}^{~A_1} (r;g) &\simeq {1 \over r^2} {g (N-1)\over 4 \pi}
\left( 1 + {g \over \pi} (N-2) \ln r \right) &\ecmb\cr
(C_m)_{\p^2 A_1}^{~\one} (r;g) &\simeq {g \over \pi r^2} {-(N-1)\over 2 N}\cr
& \quad \times
\left( 1 + {g \over \pi} \left( (N-1) a + {N-3\over 2} \ln r \right) \right)
&\ecmda a\cr
(C_m)_{\p^2 A_1}^{~A_2} (r;g) &\simeq {g \over \pi r^2} {-(N-1)\over 2}
\left( 1 + {g \over \pi} \left( - a + {2N-3 \over 2} \ln r \right) \right)
,&\ecmda b\cr}$$
where the constant $a$ is undetermined, since
\eqn\ea{A_2 - {N-1\over N}\one = {\rm O} (g) .}
The OPE coefficients $C_g$ are obtained as
\eqnn\ecgaone
\eqnn\ecgatwo
\eqna\ecgb
\eqna\ecgda
$$\eqalignno{
(C_g)_{A_1}^{~A_1} (r;g) &\simeq {N-1 \over 4 \pi r^2} \left(
1 + {g\over \pi} (N-2) \ln r \right) &\ecgaone\cr
(C_g)_{A_2}^{~A_2} (r;g) &\simeq {N \over 2 \pi r^2} \left(
1 + {g\over \pi} (N-2) \ln r \right) &\ecgatwo\cr
(C_g)_{B_0}^{~B_0} (r;g) &\simeq {-(N-2) \over 2 \pi r^2} &\ecgb a\cr
(C_g)_{B_0}^{~\one} (r;g) &\simeq - {(N-1) \over 2 \pi r^4}
\left( 1 + {g \over \pi} (N-2) (\ln r - 1) \right) &\ecgb b\cr
(C_g)_{B_0}^{~A_1} (r;g,m^2) &\simeq {m^2 \over r^2}
g {(N-1) \over 8 \pi^2} \left( - (N-1) \ln r - (N-2) \right) &\ecgb c\cr
(C_g)_{\p^2 A_1}^{~\p^2 A_1} (r;g) &\simeq {-N+3\over 4 \pi r^2} &\ecgda a\cr
(C_g)_{\p^2 A_1}^{~A_1} (r;g) &\simeq {N-1 \over \pi r^4}
\left( 1 + {g \over \pi} (N-2) (\ln r -1) \right) &\ecgda b\cr
(C_g)_{\p^2 A_1}^{~\one} (r;g,m^2) &\cr
\simeq {m^2 \over \pi r^2} &{-(N-1)(N-3) \over 4 N}
\left( 1 + {g \over \pi} (b_1 + (N-2) \ln r) \right) &\ecgda c\cr
(C_g)_{\p^2 A_1}^{~A_2} (r;g,m^2) &\cr
\simeq {m^2 \over \pi r^2} &{3(N-1)\over 4}
\left( 1 + {g \over \pi} \left( b_2 + {2 (2N-3)\over 3} \ln r \right)
\right)~,&\ecgda d\cr}$$
where the constants $b_1, b_2$ are related only by
\eqn\eb{-(N-3) b_1 + 3(N-1)b_2 = N(N-2) .}
An ambiguity is left due to Eq.~\ea.

\appendix{C}{Curvature}

At tree level we find
\eqn\eomog{\eqalign{
&\vev{\left( \O_m (r) \O_g (r') - (r \leftrightarrow r') \right) B_0
(0)}_{g,m^2}^c \cr
&\qquad\simeq \vev{ \left( \dphi (r) \phione (r') - (r \leftrightarrow r')
\right) \dphi (0)
}_{0,0}^c \cr
&\qquad = - {N-1 \over (2\pi)^3} {(r-r')_\mu\over (r-r')^2}
\left( {r_\mu \over r^2 r'^2} - 2 {r'_\mu r'_\nu r_\nu \over r'^4 r^2} +
(r \leftrightarrow r') \right) ~.}}
This has no unintegrable singularity at either $r=0$ or $r'=0$.  Hence,
Eq.~\ecurvintegral\ gives
\eqn\ecurvcal{\eqalign{
&(\Omega_{gm})_{B_0}^{~A_1} (g,m^2) \vev{A_1}_{g,m^2}\cr
&\quad \simeq - {N-1 \over (2 \pi)^3} \int_{1 \ge r} d^2 r~{\rm F.P.} \int_{1
\ge r'}
d^2 r' \cr
&\qquad\qquad \times {(r-r')_\mu \over (r-r')^2} \left(
{r_\mu\over r^2 r'^2} - 2 {r'_\mu r'_\nu r_\nu\over r^2 r'^4} + {r'_\mu \over
r^2 r'^2} - 2 {r_\mu r_\nu r'_\nu \over r^4 r'^2} \right)\cr
&\quad = - {N-1 \over (2 \pi)^3} \int_{1 \ge r} d^2 r~(- \pi) = {N-1 \over 8
\pi}~.\cr}}
The curvature for $\p^2 A_1$ can be reduced to exactly the same integral
up to a factor $-2$.

\appendix{D}{The singularities of the connection at $g=0$}

We owe the singularities of the non-maximal elements
of the connection $c_{\g}$
at $g=0$ to the singular normalization of the conjugate
field $\O_{\g}$ (see Eq.~\eogbzero).  To see this, we must go
back to the variational formula in which the connection plays
the role of finite counterterms \rsonodaiii.

Let us consider the derivative of the expectation value of
$B'_0$ in the absence of an external magnetic field.
The variational formula gives
\eqn\evarbzero{\eqalign{
- \p_g \vev{B'_0}_g &=
\int_{r \ge \ep} d^2 r~\vev{\O_{\g} (r) B'_0 (0)}_g^c \cr
&\quad + \left( (c_{\g})_{B'_0}^{~B'_0} (g) - \int_{1 \ge r \ep} {d^2 r\over
2\pi}~
(C_{\g})_{B'_0}^{~B'_0} (r;g) \right) \vev{B'_0}_g \cr
&\quad + \left( (c_{\g})_{B'_0}^{~\one} (g) - \int_{1 \ge r \ep} {d^2 r\over
2\pi}~
(C_{\g})_{B'_0}^{~\one} (r;g) \right) \cdot 1~.\cr}}
For this to be finite at $g=0$ (in fact it vanishes), the terms proportional to
${1 \over g}$ on the right-hand side must vanish.  To order ${1 \over g}$
we find
\eqn\eogbzerocor{\vev{\O_{\g} (r) B'_0 (0)}_g^c \simeq - {1 \over g}
\vev{\left[\dphi\right]_0 (r) \left[\dphi\right]_0 (0)}_0
= - {1 \over g} {N-1 \over 4 \pi^2 r^4}}
from Eq.~\ebzero.  Hence, the UV subtracted integral becomes
\eqn\evarbzeroii{\eqalign{&\int_{r \ge \ep} d^2 r~\vev{\O_{\g} (r) B'_0
(0)}_g^c
 - \int_{1 \ge r \ge \ep} {d^2 r\over 2\pi}~(C_{\g})_{B'_0}^{~\one} (r;g) \cr
&\qquad \simeq \int_{r\ge 1} d^2 r~{-1\over g} {N-1\over 4\pi^2 r^4} = - {1
\over g}
{N-1 \over 4 \pi}~.\cr}}
This singularity is canceled by the counterterm
$(c_{\g})_{B'_0}^{~\one}$ as we can see from Eq.~\enewcgb{b}.
The ${1 \over g}$ singularity of the connection is needed to cancel the
singularity in the definition of the conjugate field \eogbzero.

Similarly we can analyze the derivative of a correlation function
involving the other field ${1 \over g} \p^2 \Phi^I$:
\eqn\evarda{\eqalign{&- \p_g \vev{{1 \over g} \p^2 \Phi^I (0) \Phi^J (R)}_g
= \int_{r\ge \ep\atop |r-R|\ge \ep} d^2 r
\vev{(\O_{\g} (r) - \vev{\O_{\g}}_g) {1 \over g} \p^2 \Phi^I (0) \Phi^J
(R)}_g\cr
&\quad + \left( (c_{\g})_{{1 \over g} \p^2 \Phi^I}^{~{1 \over g} \p^2\Phi^K}
(g)
- \int_{r\ge\ep} (C_{\g})_{{1 \over g} \p^2 \Phi^I}^{~{1 \over g} \p^2 \Phi^K}
(r;g) \right)
\vev{{1 \over g} \p^2 \Phi^K (0) \Phi^J (R)}_g\cr
&\quad + \left( (c_{\g})_{{1 \over g} \p^2 \Phi^I}^{~\Phi^K} (g)
- \int_{r\ge\ep} (C_{\g})_{{1 \over g} \p^2 \Phi^I}^{~\Phi^K} (r;g) \right)
\vev{\Phi^K (0) \Phi^J (R)}_g\cr
&\quad + \left( (c_{\g})_{\Phi^J}^{~\Phi^K} (g)
- \int_{r\ge\ep} (C_{\g})_{\Phi^J}^{~\Phi^K} (r;g) \right)
\vev{{1 \over g} \p^2 \Phi^I (0) \Phi^K (R)}_g~.\cr}}
At order ${1 \over g}$, the integrand is given by
\eqn\eogdphiphi{\eqalign{
&\vev{\O_{\g} (r) {1 \over g} \p^2 \Phi^I (0)
\Phi^J (R)}_g \cr
&\quad \simeq
- {1 \over g} {\delta^{IJ}\over N} \vev{\left[ \dphi \right]_0 (r) (-2) \left[
\dphi \right]_0 (0)
\one (R)}_0 \cr
&\quad = {\delta^{IJ} \over g} {N-1\over N} {1 \over 2 \pi^2 r^4}~.\cr}}
Hence, to order ${1 \over g}$, we obtain the following UV subtracted
integral:
\eqn\evardaii{\eqalign{&\int_{r\ge \ep\atop |r-R|\ge \ep} d^2 r
\vev{(\O_{\g} (r) - \vev{\O_{\g}}_g) {1 \over g} \p^2 \Phi^I (0) \Phi^J
(R)}_g\cr
&\qquad\qquad - \int_{r\ge\ep} (C_{\g})_{{1 \over g} \p^2 \Phi^I}^{~\Phi^K}
(r;g)
\vev{\Phi^K (0) \Phi^J (R)}_g \cr
&\quad \simeq {1 \over g} \int_{r\ge 1} d^2 r {N-1\over N} {\delta^{IJ} \over 2
\pi^2 r^4}
= {1 \over g} {N-1 \over N} {\delta^{IJ} \over 2 \pi} ~.\cr}}
This is again canceled by the counterterm
\eqn\ecounter{
(c_{\g})_{{1 \over g} \p^2 \Phi^I}^{~\Phi^K} (g) \vev{\Phi^K (0) \Phi^J(0)}_g
\simeq {1 \over g} {-(N-1)\over 2\pi} {\delta^{IJ}\over N}~,}
where we have used Eq.~\enewcgda{b}.

The above cancelations which assure the validity of the variational
formula at $g=0$ can be understood as follows.  From
Eqs.~\ebzero\ and \ederivative, we observe that
the field $B'_0$ is a total derivative at lowest order in $g$:
\eqn\etotal{B'_0 \simeq \left[\dphi\right]_0 = {1 \over 2} \p^2
\left[\phione\right]_0~.}
Therefore, the UV subtracted integral
in the variational formula reduces to surface terms
at radius $|r|=1$; we have seen two examples
in the above.  To cancel these local terms
we must introduce finite counterterms
of order ${1 \over g}$.  Hence, we need the ${1\over g}$
singularities of the connection.  The only exception occurs
when the UV divergence
of the cutoff integral is logarithmic.  Since
${1 \over r^2}$ cannot be given as a total derivative,
no logarithmic divergence appears in the integral
in the variational formula, and we do not
need any subtraction or finite counterterm.
Therefore,
the maximal elements of the connection
do not need ${1 \over g}$ contributions.

Eq.~\etotal\ also explains the presence
of the extra singular factor ${1 \over g}$ in the
definition of the conjugate field, Eq.~\eogbzero.
If $B'_0$ were the conjugate field to $g$, it would
not deform the theory.
But, with ${1 \over g}$, the conjugate field is not a total derivative
at order $g^0$, and it deforms the theory non-trivially.

\appendix{E}{OPE for the derivative field $\p_\mu \Phi^N$}

Let us introduce the OPE
\eqn\eogphitoorderone{\eqalign{&2 \pi \O_g (r) \Phi^N (r') \cr
&~= (C_g)_{A_1}^{~A_1} (r-r';g) \Phi^N (r') + (C_g)_{A_1}^{~\p_\mu A_1}
(r-r';g) \p_\mu \Phi^N (r') \cr
&\qquad+ (C_g)_{A_1}^{~\p^2 A_1} (r-r';g) \p^2 \Phi^N (r') \cr
&\quad +
(C_g)_{A_1}^{~\one} (r-r';g,m^2) \one + (C_g)_{A_1}^{~A_2} (r-r';g,m^2) A_2
(r')
+ {\rm o} (|r-r'|^0)~,\cr}}
where we have ignored the term proportional to
$(2 \p_\mu \p_\nu - \delta_{\mu\nu}) A_1$, which turns out
to be irrelevant.  By taking the derivative with respect to $r'_\mu$ once
and setting $r'=0$ and taking the angular average over $r$, we obtain
\eqn\eogdphitoordertwo{2\pi \O_g (r) \p_\mu A_1 (0)
= \delta_\mu^{~\nu} (C_g)_{\p A_1}^{~\p A_1} (r;g)
\p_\nu A_1 (0) + {\rm o} \left( {1 \over r^2} \right)~,}
where we define the angular average
\eqn\edefcgdaone{\delta_\mu^{~\nu} (C_g)_{\p A_1}^{~\p A_1} (r;g)
= \delta_\mu^{~\nu} (C_g)_{A_1}^{~A_1} (r;g) -
\int_{|r|=1} {d\theta \over 2\pi}~\p_\mu (C_g)_{A_1}^{~\p_\nu A_1} (r;g)~.}
By differentiating Eq.~\eogphitoorderone\ twice with respect to $r'$
and setting $r'=0$ and taking the angular average over $r$, we obtain
\eqn\eogddphitoordertwo{\eqalign{&2 \pi \O_g (r) \p^2 A_1 (0)
= \p^2 (C_g)_{A_1}^{~A_1} (r;g) A_1 (0) + \p^2 (C_g)_{A_1}^{~\one} (r;g,m^2)
\one \cr
&~~+ \p^2 (C_g)_{A_1}^{~A_2} (r;g,m^2) A_2 (0)
+ (C_g)_{\p^2 A_1}^{~\p^2 A_1} (r;g) \p^2 A_1 (0)
+ {\rm o} \left( {1 \over r^2} \right)~,\cr}}
where
\eqn\ecgddphiresult{
(C_g)_{\p^2 A_1}^{~\p^2 A_1} (r;g) =
2 (C_g)_{\p A_1}^{~\p A_1} (r;g) - (C_g)_{A_1}^{~A_1} (r;g)~.}
Therefore, we obtain
\eqn\ecgdphiresult{
(C_g)_{\p A_1}^{~\p A_1} (r;g) =
{1 \over 2} \left( (C_g)_{A_1}^{~A_1} (r;g) + (C_g)_{\p^2 A_1}^{~\p^2 A_1}
(r;g)
\right)~.}

\listrefs
\bye